\RequirePackage{ifpdf}
\ifpdf 
\documentclass[pdftex]{sigma}
\else
\documentclass{sigma}
\fi

\numberwithin{equation}{section}
\numberwithin{theorem}{section}
\numberwithin{proposition}{section}
\numberwithin{lemma}{section}
\numberwithin{corollary}{section}
\numberwithin{definition}{section}
\numberwithin{example}{section}
\numberwithin{remark}{section}
\numberwithin{note}{section}

\begin{document}
\allowdisplaybreaks

\renewcommand{\PaperNumber}{084}

\FirstPageHeading

\ShortArticleName{Hypergeometric $\tau$ Functions of the $q$-Painlev\'e Systems of Type $(A_2+A_1)^{(1)}$}

\ArticleName{Hypergeometric $\boldsymbol{\tau}$ Functions\\ of the $\boldsymbol{q}$-Painlev\'e Systems of Type $\boldsymbol{(A_2+A_1)^{(1)}}$}

\Author{Nobutaka NAKAZONO}

\AuthorNameForHeading{N. Nakazono}

\Address{Graduate School of Mathematics, Kyushu University,
744 Motooka, Fukuoka, 819-0395, Japan}
\Email{\href{mailto:n-nakazono@math.kyushu-u.ac.jp}{n-nakazono@math.kyushu-u.ac.jp}}
\URLaddress{\url{http://researchmap.jp/nakazono/}}

\ArticleDates{Received August 17, 2010, in f\/inal form October 08, 2010;  Published online October 14, 2010}

\Abstract{We consider a $q$-Painlev\'e III equation and
a $q$-Painlev\'e II equation arising from
a birational representation of the af\/f\/ine Weyl group of type $(A_2+A_1)^{(1)}$.
We study their hypergeometric solutions
on the level of $\tau$ functions.}


\Classification{33D05; 33D15; 33E17; 39A13}

\section{Introduction}

We consider a $q$-analog of the Painlev\'e III equation ($q$-P$_{\rm III}$)
\cite{KK:qp3,KNT;projective,KNY:qp4,Sakai:Painleve}
\begin{gather}
 g_{n+1}=\cfrac{q^{2N+1}c^2}{f_ng_n}\;
  \cfrac{1+a_0q^{n}f_n}{a_0q^n+f_n},\qquad
 f_{n+1}=\cfrac{q^{2N+1}c^2}{f_ng_{n+1}}\;
  \cfrac{1+a_2a_0q^{n-m}g_{n+1}}{a_2a_0q^{n-m}+g_{n+1}},
 \label{qp3:eqn}
\end{gather}
and that of the Painlev\'e II equation ($q$-P$_{\rm II}$)
\cite{KNT;projective,RG:coales,NKT:qp2}
\begin{gather}
 X_{k+1} = \cfrac{q^{2N+1}c^2}{X_{k}X_{k-1}}\;
  \cfrac{1+a_0q^{k/2}X_k}{a_0q^{k/2}+X_k},
 \label{qp2:eqn}
\end{gather}
for the unknown functions $f_n=f_n(m,N)$, $g_n=g_n(m,N)$,
and $X_k=X_k(N)$ and the independent variables
$n,k\in\mathbb{Z}$.
Here $m,N\in\mathbb{Z}$ and $a_0,a_2,c,q\in\mathbb{C}^{\times}$
are parameters.
These equations arise from
a birational representation of the (extended) af\/f\/ine Weyl group of type $(A_2+A_1)^{(1)}$.

Note that substituting
\begin{gather*}
 m = 0,\qquad
 a_2 = q^{1/2},
\end{gather*}
and putting
\begin{gather*}
 f_k(0,N) = X_{2k}(N),\qquad
 g_k(0,N) = X_{2k-1}(N),
\end{gather*}
in (\ref{qp3:eqn}) yield (\ref{qp2:eqn}).
This procedure is called a symmetrization
of (\ref{qp3:eqn}),
which comes from the terminology used for Quispel--Roberts--Thompson (QRT) mappings~\cite{QRT1,QRT2}.

It is well known that the $\tau$ functions play a crucial
role in the theory of integrable sys\-tems~\cite{MJD:book}, and
it is also possible to introduce them in the theory
of Painlev\'e systems
\cite{JMU:Monodromy_I,JM:Monodromy_II,JM:Monodromy_III,KNY:qp4,Noumi:book,
 NY:p4,Okamoto:studies_I,Okamoto:studies_II,
 Okamoto:studies_III,Okamoto:studies_IV}.
A representation of the af\/f\/ine Weyl groups
can be lifted on the level of the $\tau$~functions
\cite{KMNOY:10E9,KMNOY:Cremona,Tsuda:tau_qp34},
which gives rise to various bilinear equations of
Hirota type satisf\/ied the $\tau$~functions.

The hypergeometric solutions of various Painlev\'e
and discrete Painlev\'e systems are expres\-sib\-le in the form of
ratio of determinants whose entries are given by hypergeometric type functions.
Usually, they are derived by reducing the bilinear equations to
the Pl\"ucker relations by using the contiguity relations satisf\/ied
by the entries of determinants
\cite{HK:a4_hyper_sol,HKW:a1+a1_hyper_sol,JKM:p4_hankel,
 KK:qp3,KM:p2_xxxiv,KNY:qp4,KO:p4_sol,KOS:dp3_casorati,KOSGR:dp2_casorati,NKT:qp2,ON:p2_sol,Sakai:qp6_sol}.
This method is elementary, but it
encounters technical dif\/f\/iculties for Painlev\'e
systems with large symmetries.
In order to overcome this dif\/f\/iculty,
Masuda has proposed a method of constructing hypergeometric solutions
under a certain boundary condition on the
lattice where the $\tau$~functions live
({\it hypergeometric $\tau$ functions}), so that they are
consistent with the action of the af\/f\/ine Weyl groups.
Although this requires somewhat complex calculations, the
merit is that it is systematic and that it can be
applied to the systems with large symmetries.
Masuda has carried out the calculations for the
$q$-Painlev\'e systems with $E_7^{(1)}$ and $E_8^{(1)}$
symmetries~\cite{Masuda:E7,Masuda:E8} and presented explicit determinant formulae for
their hypergeometric solutions.

The purpose of this paper is to apply the above method to the
$q$-Painlev\'e systems with the af\/f\/ine Weyl group symmetry of type
$(A_2+A_1)^{(1)}$ and present the explicit formulae of the
hypergeometric $\tau$ functions.
The hypergeometric $\tau$ functions provide not only determinant
formulae but also important information originating from
the geometry of lattice of the $\tau$ functions.
The result has been already announced in~\cite{KNT;projective}
and played an essential role in clarifying the mechanism of reduction from
hypergeometric solutions of (\ref{qp3:eqn}) to those of (\ref{qp2:eqn}).

This paper is organized as follows:
in Section \ref{section:qp3_and_qp2},
we f\/irst review
hypergeometric solutions of
$q$-P$_{\rm III}$ and then those of $q$-P$_{\rm II}$.
We next introduce
a representation of the af\/f\/ine Weyl group of type $(A_2+A_1)^{(1)}$.
In Section \ref{section:hyper_tau_a2a1},
we construct
the hypergeometric $\tau$ functions of $q$-P$_{\rm III}$ and those of $q$-P$_{\rm II}$.
We f\/ind that the symmetry of
the hypergeometric $\tau$ functions of $q$-P$_{\rm III}$ are connected
with Heine's transform
of the basic hypergeometric series ${}_2\varphi_1$.

We use the following conventions of $q$-analysis
throughout this paper~\cite{Gasper-Rahman:BHS}.\\
$q$-Shifted factorials:
\begin{gather*}
 (a;q)_k=\prod_{i=1}^{k}\big(1-aq^{i-1}\big).
\end{gather*}
Basic hypergeometric series:
\begin{gather*}
 {}_s\varphi_r\left(\begin{matrix}a_1,\dots,a_s\\b_1,\dots,b_r\end{matrix};q,z\right)
 =\sum_{n=0}^{\infty}\cfrac{(a_1,\dots,a_s;q)_n}{(b_1,\dots,b_r;q)_n(q;q)_n}
 \begin{bmatrix}(-1)^nq^{n(n-1)/2}\end{bmatrix} ^{1+r-s}z^n,
\end{gather*}
where
\[
(a_1,\dots,a_s;q)_n=\prod_{i=1}^s (a_i;q)_n.
\]
Jacobi theta function:
\begin{gather*}
 \Theta(a;q)=(a;q)_\infty\big(qa^{-1};q\big)_\infty.
\end{gather*}
Elliptic gamma function:
\begin{gather*}
 \Gamma(a;p,q)=\cfrac{(pqa^{-1};p,q)_{\infty}}{(a;p,q)_{\infty}},
\end{gather*}
where
\begin{gather*}
 (a;p,q)_k=\prod_{i,j=0}^{k-1} \big(1-p^iq^ja\big).
\end{gather*}
It holds that
\begin{gather*}
  \Theta(qa;q)=-a^{-1}\Theta(a;q),\qquad
 \Gamma(qa;q,q)=\Theta(a;q)\Gamma(a;q,q).
\end{gather*}

\section[$q$-P$_{\rm III}$ and $q$-P$_{\rm II}$]{$\boldsymbol{q}$-P$\boldsymbol{{}_{\rm III}}$ and $\boldsymbol{q}$-P$\boldsymbol{{}_{\rm II}}$}
\label{section:qp3_and_qp2}

\subsection[Hypergeometric solutions of $q$-P$_{\rm III}$ and $q$-P$_{\rm II}$]{Hypergeometric solutions of $\boldsymbol{q}$-P$\boldsymbol{{}_{\rm III}}$ and $\boldsymbol{q}$-P$\boldsymbol{{}_{\rm II}}$}

First, we review the hypergeometric solutions of
$q$-P$_{\rm III}$ and $q$-P$_{\rm II}$.
The hypergeometric solutions of $q$-P$_{\rm III}$ have been constructed as follows:

\begin{proposition}[\cite{KK:qp3}]
\label{prop:sol_qp3}
The hypergeometric solutions of
$q$-{\rm P}$_{\rm III}$, \eqref{qp3:eqn}, with $c=1$ are given by
\begin{gather*}
 f_n = -a_0q^n\cfrac{\psi^{n,m-1}_{N+1}\psi^{n,m}_{N}}
  {\psi^{n,m}_{N+1}\psi_{N}^{n,m-1}},\qquad
 g_n ={a_0}^{-1}a_2q^{-n-m+1}\;\cfrac{\psi^{n,m}_{N+1}\psi^{n-1,m-1}_{N}}
  {\psi^{n-1,m-1}_{N+1}\psi^{n,m}_{N}},
\end{gather*}
where $\psi^{n,m}_N$ $(N\in\mathbb{Z}_{\geq 0})$
is an $N\times N$ determinant defined by
\begin{gather*}
 \psi^{n,m}_N =
  \begin{vmatrix}
   F_{n,m}& F_{n+1,m}&\cdots&F_{n+N-1,m}\\
   F_{n-1,m}& F_{n,m}&\cdots&F_{n+N-2,m}\\
   \vdots&\vdots &\ddots&\vdots\\
   F_{n-N+1,m}&F_{n-N+2,m}&\cdots&F_{n,m}\\
  \end{vmatrix},\qquad
 \psi^{n,m}_0=1,
\end{gather*}
and $F_{n,m}$ is an arbitrary solution of the systems
\begin{gather}
F_{n+1,m} - F_{n,m} = -{a_0}^2q^{2n} F_{n,m-1},\label{qp3:contiguity_1}\\
F_{n,m+1} - F_{n,m} = -{a_2}^{-2}q^{2m+2} F_{n-1,m}.\label{qp3:contiguity_2}
\end{gather}
\end{proposition}

The general solution of
(\ref{qp3:contiguity_1}) and (\ref{qp3:contiguity_2})
 is given by
\begin{gather}
 F_{n,m}= \cfrac{A_{n,m}}{({a_2}^{-2}q^{2m+2};q^2)_\infty} \
  {}_1\varphi_1\left(\begin{matrix}0\\ {a_2}^{2}q^{-2m}\end{matrix}
  ;q^{2},{a_2}^2{a_0}^2q^{2n-2m}\right)\nonumber\\
\phantom{F_{n,m}=}{}
+B_{n,m}\cfrac{\Theta({a_0}^2{a_2}^2q^{2n-2m-2};q^2)}
  {({a_2}^2q^{-2m-2};q^2)_\infty\Theta({a_0}^2q^{2n};q^2)}~
  {}_1\varphi_1\left(\begin{matrix}0\\ {a_2}^{-2}q^{2m+4}\end{matrix}
  ;q^{2},{a_0}^2q^{2n+2}\right),\label{qp3:entry}
\end{gather}
where $A_{n,m}$ and $B_{n,m}$ are periodic functions of period one
with respect to $n$ and $m$, i.e.,
\begin{gather*}
 A_{n,m}=A_{n+1,m}=A_{n,m+1},\qquad B_{n,m}=B_{n+1,m}=B_{n,m+1}.
\end{gather*}
The explicit form of the hypergeometric solutions
of $q$-P$_{\rm II}$ are given as follows:

\begin{proposition}[\cite{NKT:qp2}]
\label{prop:sol_qp2}
The hypergeometric solutions of $q$-{\rm P}$_{\rm II}$, \eqref{qp2:eqn},
with $c=1$ are given by
\begin{gather}
 X_k = -a_0q^{k/2+N} \,\cfrac{\phi^k_{N+1}\phi^{k-1}_{N}}
 {\phi^{k-1}_{N+1}\phi^k_{N}},
\end{gather}
where $\phi^k_N$ $(N\in\mathbb{Z}_{\geq 0})$ is an $N\times N$
determinant defined by
\begin{gather}\label{qp2:det}
\phi^k_N =
 \begin{vmatrix}
  G_{k}&G_{k-1}&\cdots&G_{k-N+1}\\
  G_{k+2}&G_{k+1}&\cdots&G_{k-N+3}\\
  \vdots&\vdots&\ddots&\vdots\\
  G_{k+2N-2}&G_{k+2N-3}&\cdots&G_{k+N-1}
 \end{vmatrix},\qquad
 \phi^k_0=1,
\end{gather}
and $G_k$ is an arbitrary solution of the system
\begin{gather}\label{qp2:3-term}
 G_{k+1}-G_k+{a_0}^{-2}q^{-k}G_{k-1}=0.
\end{gather}
\end{proposition}

The general solution of (\ref{qp2:3-term}) is given by
\begin{gather}
 G_k=
 A_k\Theta(ia_0q^{(2k+1)/4};q^{1/2}) \
  {}_1\varphi_1\left(\begin{matrix}0\\-q^{1/2}\end{matrix}
  ;q^{1/2},-ia_0q^{(3+2k)/4}\right)\nonumber\\
\phantom{ G_k= }{} +B_k\Theta(-ia_0q^{(2k+1)/4};q^{1/2}) \
  {}_1\varphi_1\left(\begin{matrix}0\\-q^{1/2}\end{matrix}
  ;q^{1/2},ia_0q^{(3+2k)/4}\right),\label{qp2:entry}
\end{gather}
where $A_k$ and $B_k$ are periodic functions of period one, i.e.,
\begin{gather*}
 A_k=A_{k+1},\qquad B_k=B_{k+1}.
\end{gather*}

\subsection[Projective reduction from $q$-P$_{\rm III}$ to $q$-P$_{\rm II}$]{Projective reduction from $\boldsymbol{q}$-P$\boldsymbol{{}_{\rm III}}$ and $\boldsymbol{q}$-P$\boldsymbol{{}_{\rm II}}$}

We formulate the family of B\"acklund transformations of
$q$-P$_{\rm III}$ and $q$-P$_{\rm IV}$
as a birational representation of the
af\/f\/ine Weyl group of type $(A_2+A_1)^{(1)}$.
Here, $q$-P$_{\rm IV}$ is a $q$-analog of the Painlev\'e~IV equation
discussed in~\cite{KNY:qp4}.
We refer to \cite{Noumi:book} for basic ideas of this formulation.

We def\/ine the transformations $s_i$ ($i=0,1,2$) and $\pi$ on the variables
$f_j$ ($j=0,1,2$) and parameters $a_k$ ($k=0,1,2$) by
\begin{alignat*}{3}
& s_i(a_j) = a_j{a_i}^{-a_{ij}},\qquad
 && s_i(f_j) = f_j\left(\cfrac{a_i+f_i}{1+a_if_i}\right)^{u_{ij}},& \\
& \pi(a_i) = a_{i+1},\qquad
 &&\pi(f_i) = f_{i+1}, &
\end{alignat*}
for $i,j\in\mathbb{Z}/3\mathbb{Z}$.
Here the symmetric $3\times 3$ matrix
\begin{gather*}
 A=(a_{ij})_{i,j=0}^2
 =\left(\begin{array}{ccc}2&-1&-1\\-1&2&-1\\-1&-1&2\end{array}\right),
\end{gather*}
is the Cartan matrix of type $A_2^{(1)}$,
and the skew-symmetric one
\begin{gather*}
 U = (u_{ij})_{i,j=0}^2
 = \left(\begin{array}{ccc}0&1&-1\\-1&0&1\\1&-1&0\end{array}\right),
\end{gather*}
represents an orientation of the corresponding Dynkin diagram.
We also def\/ine the transformations $w_j$ ($j=0,1$) and $r$ by
\begin{alignat*}{3}
&  w_0(f_i)=\cfrac{a_ia_{i+1}(a_{i-1}a_i+a_{i-1}f_i+f_{i-1}f_i)}
  {f_{i-1}(a_ia_{i+1}+a_if_{i+1}+f_if_{i+1})},\qquad
 &&w_0(a_i)=a_i,&\\
&  w_1(f_i)=\cfrac{1+a_if_i+a_ia_{i+1}f_if_{i+1}}
  {a_ia_{i+1}f_{i+1}(1+a_{i-1}f_{i-1}+a_{i-1}a_if_{i-1}f_i)},\qquad
 &&w_1(a_i)=a_i,&\\
& r(f_i)=\cfrac{1}{f_i},\qquad
 &&r(a_i)=a_i,&
\end{alignat*}
for $i\in\mathbb{Z}/3\mathbb{Z}$.

\begin{proposition}[\cite{KNY:qp4}]
The group of birational transformations
$\langle s_0,s_1,s_2,\pi, w_0,w_1,r\rangle$
forms the affine Weyl group of type $(A_2+A_1)^{(1)}$,
denoted by $\widetilde{W}((A_2+A_1)^{(1)})$.
Namely, the transformations satisfy the fundamental relations
\begin{gather*}
 {s_i}^2=(s_is_{i+1})^3=\pi^3=1,\qquad
 \pi s_i = s_{i+1}\pi\qquad
 (i\in\mathbb{Z}/3\mathbb{Z}),\\
 {w_0}^2={w_1}^2=r^2=1,\qquad
 rw_0=w_1r,
\end{gather*}
and the action of $\widetilde{W}(A_2^{(1)})=\langle s_0,s_1,s_2,\pi\rangle$ and
that of $\widetilde{W}(A_1^{(1)})=\langle w_0,w_1,r\rangle$ commute with each other.
\end{proposition}

In general, for a function $F=F(a_i,f_j)$, we let an element
$w\in\widetilde{W}((A_2+A_1)^{(1)})$
act as $w.F(a_i,f_j)=F(a_i.w,f_j.w)$, that is, $w$
acts on the arguments from the right.  Note that $a_0a_1a_2=q$
and $f_0f_1f_2=qc^2$ are invariant under the action of
$\widetilde{W}((A_2+A_1)^{(1)})$
and $\widetilde{W}(A_2^{(1)})$,
respectively. We def\/ine the translations $T_i$ ($i=1,2,3,4$) by
\begin{gather}\label{eqn:translations}
 T_1=\pi s_2s_1,\qquad
 T_2=s_1\pi s_2,\qquad
 T_3=s_2s_1\pi,\qquad
 T_4=rw_0,
\end{gather}
whose action on parameters $a_i$ $(i=0,1,2)$ and $c$ is given by
\begin{gather*}
  T_1:~(a_0,a_1,a_2,c)\mapsto\big(qa_0,q^{-1}a_1,a_2,c\big),\\
  T_2:~(a_0,a_1,a_2,c)\mapsto(a_0,qa_1,q^{-1}a_2,c),\\
  T_3:~(a_0,a_1,a_2,c)\mapsto\big(q^{-1}a_0,a_1,qa_2,c\big),\\
  T_4:~(a_0,a_1,a_2,c)\mapsto(a_0,a_1,a_2,qc).
\end{gather*}
Note that $T_i$ ($i=1,2,3,4$) commute with each other and $T_1T_2T_3=1$.
The action of $T_1$ on the $f$-variables
can be expressed as
\begin{gather}\label{qp3:eqn2}
 T_1(f_1) = \cfrac{qc^2}{f_1f_0} \cfrac{1+a_0f_0}{a_0+f_0},\qquad
 T_1(f_0) = \cfrac{qc^2}{f_0T_1(f_1)} \cfrac{1+a_2a_0T_1(f_1)}{a_2a_0+T_1(f_1)}.
\end{gather}
Or, applying ${T_1}^n{T_2}^m{T_4}^N$ $(n,m,N\in\mathbb{Z})$
on (\ref{qp3:eqn2}) and putting
\begin{gather*}
 f_{i,N}^{n,m} = {T_1}^n{T_2}^m{T_4}^N(f_i)\qquad (i=0,1,2),
\end{gather*}
we obtain
\begin{gather*}
 f_{1,N}^{n+1,m} = \cfrac{q^{2N+1}c^2}{f_{1,N}^{n,m}f_{0,N}^{n,m}}\;
   \cfrac{1+a_0q^nf_{0,N}^{n,m}}{a_0q^n+f_{0,N}^{n,m}},\qquad
 f_{0,N}^{n+1,m} = \cfrac{q^{2N+1}c^2}{f_{0,N}^{n,m}f_{1,N}^{n+1,m}}\;
   \cfrac{1+a_2a_0q^{n-m}f_{1,N}^{n+1,m}}{a_2a_0q^{n-m}+f_{1,N}^{n+1,m}},
\end{gather*}
which is equivalent to $q$-P$_{\rm III}$.
Then $T_1$ and $T_i$ ($i=2,4$) are regarded as the time evolution and
B\"acklund transformations of $q$-P$_{\rm III}$, respectively.
We here note that we also obtain $q$-P$_{\rm IV}$ by identifying
$T_4$ as a time evolution~\cite{KNY:qp4}.

In order to formulate the symmetrization to $q$-P$_{\rm II}$,
it is crucial to introduce the transforma\-tion~$R_1$ def\/ined by
\begin{gather}\label{eqn:def_R1}
 R_1 = \pi^2 s_1,
\end{gather}
which satisf\/ies
\begin{gather*}
 {R_1}^2 = T_1.
\end{gather*}
Considering the projection of the action of $R_1$
on the line $a_2=q^{1/2}$, we have
\begin{gather}
  R_1:~(a_0,a_1,c)\mapsto(q^{1/2}a_0,q^{-1/2}a_1,c),\nonumber\\
  R_1(f_0) = \cfrac{qc^2}{f_0f_1}\;\cfrac{1+a_0f_0}{a_0+f_0},\qquad
 R_1(f_1) = f_0.\label{qp2:eqn2}
\end{gather}
Applying ${R_1}^k{T_4}^N$ on (\ref{qp2:eqn2}) and putting
\begin{gather*}
 f_{i,N}^{k} = {R_1}^k{T_4}^N(f_i)\qquad (i=0,1,2),
\end{gather*}
we have
\begin{gather*}
 f_{0,N}^{k+1} = \cfrac{q^{2N+1}c^2}{f_{0,N}^{k} f_{0,N}^{k-1}}\;
 \cfrac{1+a_0q^{k/2} f_{0,N}^{k}}{a_0q^{k/2} + f_{0,N}^{k}},
\end{gather*}
which is equivalent to $q$-P$_{\rm II}$.
Then $R_1$ and $T_4$ are regarded as
the time evolution and a B\"acklund transformation
of $q$-P$_{\rm II}$, respectively.

In general, we can derive various discrete
Painlev\'e systems from elements of inf\/inite order of af\/f\/ine Weyl groups
that are not necessarily translations by taking a
projection on a certain subspace of the parameter space.
We call such a procedure a projective reduction~\cite{KNT;projective}.
The symmetrization is a kind of the projective reduction.

\subsection[Birational representation of $\widetilde{W}((A_2+A_1)^{(1)})$ on the $\tau$ function]{Birational representation of $\boldsymbol{\widetilde{W}((A_2+A_1)^{(1)})}$ on the $\boldsymbol{\tau}$ function}

We introduce the new variables $\tau_i$ and
$\overline{\tau}_i$ $(i\in \mathbb{Z}/3\mathbb{Z})$
by letting
\begin{gather*}
 f_i=q^{1/3}c^{2/3}\,\cfrac{\overline{\tau}_{i+1}\tau_{i-1}}
  {\tau_{i+1}\overline{\tau}_{i-1}},
\end{gather*}
and lift
a representation to the af\/f\/ine Weyl group
on their level:

\begin{proposition}[\cite{Tsuda:tau_qp34}]\label{prop:weyl_group_tau}
We define the action of $s_i$ $(i=0,1,2)$,
$\pi$, $w_j$ $(j=0,1)$, and $r$ on $\tau_k$ and
$\overline{\tau}_k$ $(k=0,1,2)$ by the following formulae:
\begin{alignat*}{3}
& s_i(\tau_i)=
  \cfrac{u_i\tau_{i+1}\overline{\tau}_{i-1}+\overline{\tau}_{i+1}\tau_{i-1}}
   {{u_i}^{1/2}\overline{\tau}_i},
 && s_i(\tau_j) = \tau_j\ \; (i\neq j),& \\
& s_i(\overline{\tau}_i)=
  \cfrac{v_i\overline{\tau}_{i+1}\tau_{i-1}+\tau_{i+1}\overline{\tau}_{i-1}}
   {{v_i}^{1/2}\tau_i},
 &&s_i(\overline{\tau}_j) = \overline{\tau}_j\ \; (i\neq j), &\\
&
 \pi(\tau_i)= \tau_{i+1},
 &&\pi(\overline{\tau}_i) = \overline{\tau}_{i+1},& \\
&
 w_0(\overline{\tau}_i)=
  \cfrac{{a_{i+1}}^{1/3}(\overline{\tau}_i\tau_{i+1}\tau_{i+2}\!
   + u_{i-1}\tau_i\overline{\tau}_{i+1}\tau_{i+2}\!
   + {u_{i+1}}^{-1}\tau_i\tau_{i+1}\overline{\tau}_{i+2})}
  {{a_{i+2}}^{1/3}\overline{\tau}_{i+1}\overline{\tau}_{i+2}},
\qquad &&w_0(\tau_i) = \tau_i,&\\
&
 w_1(\tau_i)=
  \cfrac{{a_{i+1}}^{1/3}(\tau_i\overline{\tau}_{i+1}\overline{\tau}_{i+2}\!
   + v_{i-1}\overline{\tau}_i\tau_{i+1}\overline{\tau}_{i+2}\!
   + {v_{i+1}}^{-1}\overline{\tau}_i\overline{\tau}_{i+1}\tau_{i+2})}
  {{a_{i+2}}^{1/3}\tau_{i+1}\tau_{i+2}},
 &&w_1(\overline{\tau}_i) = \overline{\tau}_i, &\\
& r(\tau_i)= \overline{\tau}_i,
 &&r(\overline{\tau}_i) = \tau_i,&
\end{alignat*}
with
\begin{gather*}
 u_i = q^{-1/3}c^{-2/3}a_i,\qquad
 v_i = q^{1/3}c^{2/3}a_i,
\end{gather*}
where $i,j\in\mathbb{Z}/3\mathbb{Z}$.
Then, $\langle s_0,s_1,s_2,\pi,w_0,w_1,r\rangle$ forms
the affine Weyl group $\widetilde{W}((A_2+A_1)^{(1)})$.
\end{proposition}

We def\/ine the $\tau$ function $\tau^{n,m}_{N}$
($n,m,N\in\mathbb{Z}$) by
\begin{gather*}
 \tau^{n,m}_{N}={T_1}^n{T_2}^m{T_4}^N(\tau_1).
\end{gather*}
We note that
\begin{gather}\label{eqn:taunmN_tau}
 \tau_0=\tau^{-1,0}_{0},\qquad\!
 \tau_1=\tau^{0,0}_{0},\qquad\!
 \tau_2=\tau^{0,1}_{0},\qquad\!
 \overline{\tau}_0=\tau^{-1,0}_{1},\qquad\!
 \overline{\tau}_1=\tau^{0,0}_{1},\qquad\!
 \overline{\tau}_2=\tau^{0,1}_{1},
\end{gather}
and
\begin{gather*}
 f_{0,N}^{n,m} = q^{(2N+1)/3}c^{2/3}\,
  \cfrac{\tau^{n,m}_{N+1}\tau^{n,m+1}_{N}}
  {\tau_N^{n,m}\tau^{n,m+1}_{N+1}},\qquad
 f_{1,N}^{n,m} = q^{(2N+1)/3}c^{2/3}\,
  \cfrac{\tau^{n,m+1}_{N+1}\tau^{n-1,m}_{N}}
  {\tau_N^{n,m+1}\tau^{n-1,m}_{N+1}},\\
 f_{2,N}^{n,m} = q^{(2N+1)/3}c^{2/3}\,
  \cfrac{\tau^{n-1,m}_{N+1}\tau^{n,m}_{N}}
  {\tau_N^{n-1,m}\tau^{n,m}_{N+1}}.
\end{gather*}
Let us consider the $\tau$ functions for $q$-P$_{\rm II}$.
We set
\begin{gather*}
 \tau^k_N = {R_1}^k{T_4}^N(\tau_1).
\end{gather*}
Note that
\begin{gather}\label{eqn:taukN_tau}
 \tau_0 = \tau^{-2}_0,\qquad
 \tau_1 = \tau^{0}_0,\qquad
 \tau_2 = \tau^{-1}_0,\qquad
 \overline{\tau}_0 = \tau^{-2}_1,\qquad
 \overline{\tau}_1 = \tau^{0}_1,\qquad
 \overline{\tau}_2 = \tau^{-1}_1,
\end{gather}
and
\begin{gather*}
 f_{0,N}^{k}=q^{(2N+1)/3}c^{2/3}\,
 \cfrac{\tau^{k}_{N+1}\tau^{k-1}_N}{\tau^k_N\tau^{k-1}_{N+1}}.
\end{gather*}
In general, it follows that
\begin{gather*}
 \tau^{n,0}_N = \tau^{2n}_{N},\qquad  \tau^{n,1}_N = \tau^{2n-1}_{N}.
\end{gather*}
For convenience, we introduce $\alpha_i$, $\gamma$, and $Q$ by
\begin{gather*}
 {\alpha_i}^6 = a_i,\qquad \gamma^6 = c,\qquad Q^6=q.
\end{gather*}

\section[Hypergeometric $\tau$ functions of the $q$-Painlev\'e systems of type $(A_2+A_1)^{(1)}$]{Hypergeometric $\boldsymbol{\tau}$ functions of the $\boldsymbol{q}$-Painlev\'e systems\\ of type $\boldsymbol{(A_2+A_1)^{(1)}}$}
\label{section:hyper_tau_a2a1}

In this section, we construct the hypergeometric $\tau$ functions
of $q$-P$_{\rm III}$ and $q$-P$_{\rm II}$.
We def\/ine the hypergeometric $\tau$ functions of $q$-P$_{\rm III}$
by $\tau_N^{n,m}$ consistent with the action of $\langle T_1,T_2,T_3,T_4\rangle$.
We also def\/ine the hypergeometric $\tau$ functions of $q$-P$_{\rm II}$
by $\tau_N^k$ consistent with the action of $\langle R_1,T_4\rangle$.
Here, we mean $\tau(\alpha)$ consistent with a action of transformation $r$ as
\begin{gather*}
 r.\tau(\alpha)=\tau(\alpha.r).
\end{gather*}
We then regard $\tau_N^{n,m}$
as function in $\alpha_0$ and $\alpha_2$, i.e.,
\begin{gather*}
 \tau_N^{n,m}=\tau_N^{0,0}(Q^n\alpha_0,Q^{-m}\alpha_2).
\end{gather*}
We also regard $\tau_N^k$ as function in $\alpha_0$, i.e.,
\begin{gather*}
 \tau_N^k=\tau_N^0(Q^{k/2}\alpha_0).
\end{gather*}

\subsection[Hypergeometric $\tau$ functions of $q$-P$_{\rm III}$]{Hypergeometric $\boldsymbol{\tau}$ functions of $\boldsymbol{q}$-P$\boldsymbol{{}_{\rm III}}$}
\label{section:hyper_tau_qp3}

We construct the hypergeometric $\tau$ functions of $q$-P$_{\rm III}$.
By the action of the af\/f\/ine Weyl group,
$\tau_N^{n,m}$ is determined as a rational function in
$\tau_0^{n,m}$ and $\tau_1^{n,m}$ (or $\tau_i$ and $\overline{\tau}_i$).
Thus, our purpose is determining $\tau_0^{n,m}$ and $\tau_1^{n,m}$
consistent with the
action of $\langle T_1,T_2,T_3,T_4\rangle$ and constructing $\tau_N^{n,m}$
under the condition
\begin{gather}\label{eqn:qp3_condition_gamma}
 \gamma=1,
\end{gather}
and the boundary condition
\begin{gather}\label{eqn:qp3_condition_tau-1}
 \tau_N^{n,m}=0\quad (N<0).
\end{gather}

First we consider the condition for $\tau_0^{n,m}$
which follows from the boundary condition~(\ref{eqn:qp3_condition_tau-1}).
We use the bilinear equations obtained in~\cite{KNT;projective}:

\begin{proposition}
The following bilinear equations hold:
\begin{gather}
  \tau^{n,m}_{N+1}\tau^{n,m}_{N-1}
  + Q^{4n-8m+4}{\alpha_1}^{-4}{\alpha_2}^{4}\left(\tau^{n,m}_{N}\right)^2
  - Q^{n-2m+1}{\alpha_1}^{-1}\alpha_2\tau^{n,m+1}_{N}\tau^{n,m-1}_{N} = 0,
 \label{eqn:TypeI_1}\\
  \tau^{n,m}_{N+1}\tau^{n,m}_{N-1}
  + Q^{4n+4m}{\alpha_0}^{4}{\alpha_2}^{-4}\left(\tau^{n,m}_{N}\right)^2
  - Q^{n+m}\alpha_0{\alpha_2}^{-1}\tau^{n+1,m+1}_{N}\tau^{n-1,m-1}_{N} = 0,\\
  \tau^{n,m}_{N+1}\tau^{n,m}_{N-1}
  + Q^{-8n+4m-4}{\alpha_0}^{-4}{\alpha_1}^{4}\left(\tau^{n,m}_{N}\right)^2
  - Q^{-2n+m-1}{\alpha_0}^{-1}\alpha_1\tau^{n+1,m}_{N}\tau^{n-1,m}_{N} = 0.
 \label{eqn:TypeI_3}
\end{gather}
\end{proposition}

By putting $N=0$ in (\ref{eqn:TypeI_1})--(\ref{eqn:TypeI_3}),
we get
\begin{gather}
  Q^{4n-8m+4}{\alpha_1}^{-4}{\alpha_2}^{4}\left(\tau^{n,m}_{0}\right)^2
  - Q^{n-2m+1}{\alpha_1}^{-1}\alpha_2\tau^{n,m+1}_{0}\tau^{n,m-1}_{0} = 0,
 \label{eqn:TypeI_1_2}\\
  Q^{4n+4m}{\alpha_0}^{4}{\alpha_2}^{-4}\left(\tau^{n,m}_{0}\right)^2
  - Q^{n+m}\alpha_0{\alpha_2}^{-1}\tau^{n+1,m+1}_{0}\tau^{n-1,m-1}_{0} = 0,\\
  Q^{-8n+4m-4}{\alpha_0}^{-4}{\alpha_1}^{4}\left(\tau^{n,m}_{0}\right)^2
  - Q^{-2n+m-1}{\alpha_0}^{-1}\alpha_1\tau^{n+1,m}_{0}\tau^{n-1,m}_{0} = 0.
 \label{eqn:TypeI_3_2}
\end{gather}
We set
\begin{gather}\label{eqn:gage_qp3_tau0}
 \tau^{n,m}_{0} =
 \Gamma(Q^{2n-m+1}{\alpha_0}^{2}\alpha_2;Q,Q)
 \Gamma(Q^{-n+2m-1}{\alpha_1}^{2}\alpha_0;Q,Q)
 \Gamma(Q^{-n-m}{\alpha_2}^{2}\alpha_1;Q,Q) A^{n,m}_{0}.\!\!\!\!
\end{gather}
From \eqref{eqn:TypeI_1_2}--\eqref{eqn:TypeI_3_2},
the following equations hold:
\begin{gather}
(A^{n,m}_{0})^2 = A^{n,m+1}_{0}A^{n,m-1}_{0},\label{eqn:tau_0_1}\\
(A^{n,m}_{0})^2 = A^{n+1,m+1}_{0}A^{n-1,m-1}_{0},\\
(A^{n,m}_{0})^2 = A^{n+1,m}_{0}A^{n-1,m}_{0}.
\end{gather}

We next determine $\tau_0^{n,m}$ and $\tau_1^{n,m}$.
From (\ref{eqn:translations}) and Proposition \ref{prop:weyl_group_tau},
we see that the action of~$T_1$,~$T_2$, and $T_3$ are given by
\begin{gather}
 T_{i}(\tau_{i-1}) = \tau_{i},\label{eqn:para_tau_1}\\
 T_{i}(\overline{\tau}_{i-1}) = \overline{\tau}_{i},\label{eqn:para_tau_2}\\
 T_{i}(\tau_{i+1}) =
 \cfrac{{\alpha_{i-1}}^6\tau_{i}\overline{\tau}_{i+1}+Q^2\overline{\tau}_{i}\tau_{i+1}}
  {Q{\alpha_{i-1}}^3\overline{\tau}_{i-1}},\label{eqn:para_tau_3}\\
 T_{i}(\overline{\tau}_{i+1}) =
 \cfrac{Q^2{\alpha_{i-1}}^6\tau_{i+1}\overline{\tau}_{i}+\overline{\tau}_{i+1}\tau_{i}}
  {Q{\alpha_{i-1}}^3\tau_{i-1}},\label{eqn:para_tau_4}\\
 T_{i}(\tau_{i}) =
 \cfrac{1}{{\alpha_{i+1}}^{3}{\alpha_{i-1}}^{6}}\; \cfrac{{\tau_{i}}^2}{\tau_{i-1}}
 +\cfrac{\alpha_{i+1}{\alpha_{i-1}}^{4}}{{\alpha_{i}}^{2}}\;
  \cfrac{\tau_{i}\overline{\tau}_{i}}{\overline{\tau}_{i-1}}
 +\cfrac{{\alpha_{i}}^{2}{\alpha_{i-1}}^{2}}{\alpha_{i+1}}\;
  \cfrac{\overline{\tau}_{i}\tau_{i}\tau_{i+1}}{\overline{\tau}_{i+1}\tau_{i-1}}
 +{\alpha_{i+1}}^{3}\,
  \cfrac{{\overline{\tau}_{i}}^2\tau_{i+1}}
  {\overline{\tau}_{i+1}\overline{\tau}_{i-1}},\label{eqn:para_tau_5}\\
 T_{i}(\overline{\tau}_{i}) =
  \cfrac{1}{{\alpha_{i+1}}^{3}{\alpha_{i-1}}^{6}}\;
  \cfrac{{\overline{\tau}_{i}}^2}{\overline{\tau}_{i-1}}
 +{\alpha_{i}}^{2}{\alpha_{i+1}}^{5}{\alpha_{i-1}}^{8}
  \cfrac{\tau_{i}\overline{\tau}_{i}}{\tau_{i-1}}\nonumber\\
 \phantom{T_{i}(\overline{\tau}_{i}) =}{}
 +\cfrac{1}{{\alpha_{i}}^{2}{\alpha_{i+1}}^{5}{\alpha_{i-1}}^{2}}\;
  \cfrac{\tau_{i}\overline{\tau}_{i}\overline{\tau}_{i+1}}{\tau_{i+1}\overline{\tau}_{i-1}}
 +{\alpha_{i+1}}^{3}\,\cfrac{{\tau_{i}}^2\overline{\tau}_{i+1}}
  {\tau_{i+1}\tau_{i-1}},\label{eqn:para_tau_6}
\end{gather}
where $i=1,2,3$.

\begin{lemma}
If $\tau_i$ and $\overline{\tau}_i$
are consistent with
\eqref{eqn:para_tau_1}--\eqref{eqn:para_tau_4},
then they are also consistent with~\eqref{eqn:para_tau_5} and \eqref{eqn:para_tau_6}.
\end{lemma}

\begin{proof}
Applying $T_{i-1}$ on (\ref{eqn:para_tau_4}) and using
(\ref{eqn:para_tau_1}) and (\ref{eqn:para_tau_2}),
we have
\begin{gather}\label{eqn:Ti_taui}
 T_{i}(\tau_{i}) =
 \cfrac{Q}{{\alpha_{i+1}}^{3}{\alpha_{i-1}}^{3}}\;
  \cfrac{\tau_{i}}{\overline{\tau}_{i+1}}\,T_{i}(\overline{\tau}_{i+1})
 + \cfrac{{\alpha_{i+1}}^2{\alpha_{i-1}}^2}{\alpha_{i}}\;
  \cfrac{\overline{\tau}_{i}}{\overline{\tau}_{i+1}}\,T_{i}(\tau_{i+1}).
\end{gather}
By using  (\ref{eqn:para_tau_3}) and (\ref{eqn:para_tau_4})
for (\ref{eqn:Ti_taui}), we get (\ref{eqn:para_tau_5}).
Similarly, applying $T_{i-1}$ on (\ref{eqn:para_tau_3})
and using~(\ref{eqn:para_tau_1}) and~(\ref{eqn:para_tau_2}),
we have
\begin{gather}\label{eqn:Ti_overtaui}
 T_{i}(\overline{\tau}_{i}) =
  \cfrac{1}{Q{\alpha_{i+1}}^{3}{\alpha_{i-1}}^{3}}\;
   \cfrac{\overline{\tau}_{i}}{\tau_{i+1}}\,T_{i}(\tau_{i+1})
  + Q{\alpha_{i-1}}^{3}{\alpha_{i+1}}^{3}\,
   \cfrac{\tau_{i}}{\tau_{i+1}}\,T_{i}(\overline{\tau}_{i+1}).
\end{gather}
By using (\ref{eqn:para_tau_3}) and (\ref{eqn:para_tau_4})
for (\ref{eqn:Ti_overtaui}), we get (\ref{eqn:para_tau_6}).
\end{proof}

From (\ref{eqn:taunmN_tau}), we rewrite (\ref{eqn:para_tau_3})
and (\ref{eqn:para_tau_4}) as follows:
\begin{gather}
 \tau_{0}^{-1,0}\tau_{1}^{0,1}
 - Q^{-1}{\alpha_{1}}^3\tau_{0}^{-1,1}\tau_{1}^{0,0}
 + Q^{-2}{\alpha_{1}}^6\tau_{0}^{0,1}\tau_{1}^{-1,0}
 = 0,\label{eqn:para_tau_3_1}\\
 \tau_{0}^{0,0}\tau_{1}^{-1,0}
 - Q^{-1}{\alpha_{2}}^3\tau_{0}^{-1,-1}\tau_{1}^{0,1}
 + Q^{-2}{\alpha_{2}}^6\tau_{0}^{-1,0}\tau_{1}^{0,0}
 = 0,\\
 \tau_{0}^{0,1}\tau_{1}^{0,0}
 - Q^{-1}{\alpha_{0}}^3\tau_{0}^{1,1}\tau_{1}^{-1,0}
 + Q^{-2}{\alpha_{0}}^6\tau_{0}^{0,0}\tau_{1}^{0,1}
 = 0,\label{eqn:para_tau_3_3}\\
 \tau_{0}^{0,1}\tau_{1}^{-1,0}
 - Q{\alpha_{1}}^3\tau_{0}^{0,0}\tau_{1}^{-1,1}
 + Q^2{\alpha_{1}}^6\tau_{0}^{-1,0}\tau_{1}^{0,1}
 = 0,\label{eqn:para_tau_4_1}\\
 \tau_{0}^{-1,0}\tau_{1}^{0,0}
 - Q{\alpha_{2}}^3\tau_{0}^{0,1}\tau_{1}^{-1,-1}
 + Q^2{\alpha_{2}}^6\tau_{0}^{0,0}\tau_{1}^{-1,0}
 = 0,\\
 \tau_{0}^{0,0}\tau_{1}^{0,1}
 - Q{\alpha_{0}}^3\tau_{0}^{-1,0}\tau_{1}^{1,1}
 + Q^2{\alpha_{0}}^6\tau_{0}^{0,1}\tau_{1}^{0,0}
 = 0.\label{eqn:para_tau_4_3}
\end{gather}
We set
\begin{gather}\label{eqn:gage_qp3_tau1}
 \tau^{n,m}_{1} =
 -Q^{2n+2m}{\alpha_0}^{2}{\alpha_2}^{-2}\,
  \cfrac{\Theta(-Q^{-6n}{\alpha_0}^{-6};Q^6)
   \Theta(-Q^{6m}{\alpha_2}^{-6};Q^6)}
   {\Theta(Q^{-6(n-m)}{\alpha_0}^{-6}{\alpha_2}^{-6};Q^6)}\,
  \tau_0^{n,m}F_{n,m-1}.
\end{gather}
Here, $F_{n,m}$ is equivalent to (\ref{qp3:entry})
because we obtain
(\ref{qp3:contiguity_1}) and (\ref{qp3:contiguity_2})
from (\ref{eqn:para_tau_4_1})--(\ref{eqn:para_tau_4_3}) and
(\ref{eqn:para_tau_3_1})--(\ref{eqn:para_tau_3_3}), respectively.
If we assume $A_0^{n,m}$ is an arbitrary constant,
it does not contradict (\ref{eqn:tau_0_1})--(\ref{eqn:para_tau_6}).
Therefore, we may set $A_0^{n,m}=1$.

Finally we construct $\tau_N^{n,m}$.
\begin{theorem}
Under the assumption \eqref{eqn:qp3_condition_gamma} and
\eqref{eqn:qp3_condition_tau-1},
the hypergeometric $\tau$ functions of $q$-{\rm P}$_{\rm III}$ are given
as the follows:
\begin{gather}
 \tau^{n,m}_N =
  (-1)^{N(N+1)/2}Q^{-2(2n-m)N^2+6nN}{\alpha_0}^{-4N^2+6N}{\alpha_2}^{-2N^2}
  \nonumber\\
\phantom{\tau^{n,m}_N =}{}
 \times
  \left(\cfrac{\Theta(-Q^{-6n}{\alpha_0}^{-6};Q^6)
   \Theta(-Q^{6m}{\alpha_2}^{-6};Q^6)}
   {\Theta(Q^{-6(n-m)}{\alpha_0}^{-6}{\alpha_2}^{-6};Q^6)}
  \right)^N\nonumber\\
\phantom{\tau^{n,m}_N =}{}
  \times\Gamma(Q^{2n-m+1}{\alpha_0}^{2}\alpha_2;Q,Q)
   \Gamma(Q^{-n+2m-1}{\alpha_1}^{2}\alpha_0;Q,Q)\nonumber\\
\phantom{\tau^{n,m}_N =}{} \times
   \Gamma(Q^{-n-m}{\alpha_2}^{2}\alpha_1;Q,Q) \psi^{n,m-1}_{N},
  \label{eqn:qp3_hyper_tau}
\end{gather}
where
\begin{gather*}
 \psi_N^{n,m} =
 \begin{vmatrix}
   F_{n,m}& F_{n+1,m}&\cdots&F_{n+N-1,m}\\
   F_{n-1,m}& F_{n,m}&\cdots&F_{n+N-2,m}\\
   \vdots&\vdots &\ddots&\vdots\\
   F_{n-N+1,m}&F_{n-N+2,m}&\cdots&F_{n,m}\\
 \end{vmatrix},\qquad
 \psi_0^{n,m}=1,\qquad \psi_{-N}^{n,m}=0 \quad (N>0),
\end{gather*}
and
\begin{gather}
 F_{n,m}= \cfrac{A_{n,m}}{({a_2}^{-2}q^{2m+2};q^2)_\infty} \
  {}_1\varphi_1\left(\begin{matrix}0\\ {a_2}^{2}q^{-2m}\end{matrix}
  ;q^{2},{a_2}^2{a_0}^2q^{2n-2m}\right)\nonumber\\
  \phantom{F_{n,m}=}{}
  +B_{n,m}\cfrac{\Theta({a_0}^2{a_2}^2q^{2n-2m-2};q^2)}
  {({a_2}^2q^{-2m-2};q^2)_\infty\Theta({a_0}^2q^{2n};q^2)} \
  {}_1\varphi_1\left(\begin{matrix}0\\ {a_2}^{-2}q^{2m+4}\end{matrix}
  ;q^{2},{a_0}^2q^{2n+2}\right).\label{qp3:entry_2}
\end{gather}
Here, $A_{n,m}$ and $B_{n,m}$ are periodic functions of period one
with respect to $n$ and $m$.
\end{theorem}

\begin{proof}
We set
\begin{gather*}
 \tau^{n,m}_N  =
  (-1)^{N(N+1)/2}Q^{-2(2n-m)N^2+6nN}{\alpha_0}^{-4N^2+6N}{\alpha_2}^{-2N^2}\\
  \phantom{\tau^{n,m}_N  =}{}\times
  \left(\cfrac{\Theta(-Q^{-6n}{\alpha_0}^{-6};Q^6)
   \Theta(-Q^{6m}{\alpha_2}^{-6};Q^6)}
   {\Theta(Q^{-6(n-m)}{\alpha_0}^{-6}{\alpha_2}^{-6};Q^6)}
  \right)^N\nonumber\\
  \phantom{\tau^{n,m}_N  =}{}
   \times\Gamma(Q^{2n-m+1}{\alpha_0}^{2}\alpha_2;Q,Q)
   \Gamma(Q^{-n+2m-1}{\alpha_1}^{2}\alpha_0;Q,Q)
   \Gamma(Q^{-n-m}{\alpha_2}^{2}\alpha_1;Q,Q) \psi^{n,m-1}_{N}.
\end{gather*}
From (\ref{eqn:qp3_condition_tau-1}), (\ref{eqn:gage_qp3_tau0}),
and (\ref{eqn:gage_qp3_tau1}),
we f\/ind
\begin{gather*}
 \psi_N^{n,m}=0 \quad (N<0),\qquad
 \psi_0^{n,m}=1,\qquad
 \psi_1^{n,m}=F_{n,m}.
\end{gather*}
Furthermore, it is easily verif\/ied that $\psi_N^{n,m}$ satisfy
\begin{gather}\label{eqn:qp3_toda_bilinear}
 \psi^{n,m}_{N+1}\psi^{n,m}_{N-1}
 -\left(\psi^{n,m}_{N}\right)^2
 +\psi^{n+1,m}_{N}\psi^{n-1,m}_{N} = 0,
\end{gather}
from $(\ref{eqn:TypeI_3})$.
In general, (\ref{eqn:qp3_toda_bilinear}) admits a solution expressed
in terms of the Toeplitz type determinant
\begin{gather*}
 \psi^{n,m}_{N} = \det\left(c_{n-i+j,m}\right)_{i,j=1,\ldots,N}\quad (N>0),
\end{gather*}
under the boundary conditions
\begin{gather*}
 \psi^{n,m}_{N} = 0\quad (N<0),\qquad
 \psi^{n,m}_0 = 1,\qquad
 \psi^{n,m}_1 = c_{n,m},
\end{gather*}
where $c_{n,m}$ is an arbitrary function.
Therefore we have completed the proof.
\end{proof}

\subsection[Hypergeometric $\tau$ functions of $q$-P$_{\rm II}$]{Hypergeometric $\boldsymbol{\tau}$ functions of $\boldsymbol{q}$-P$\boldsymbol{{}_{\rm II}}$}

In this section,
we construct the hypergeometric $\tau$ functions of $q$-P$_{\rm II}$
by two methods.

\subsubsection[Hypergeometric $\tau$ functions of $q$-P$_{\rm II}$ (I)]{Hypergeometric $\boldsymbol{\tau}$ functions of $\boldsymbol{q}$-P$\boldsymbol{{}_{\rm II}}$ (I)}

We construct the hypergeometric $\tau$ functions of $q$-P$_{\rm II}$
by using those of $q$-P$_{\rm III}$.
We here note that
$\tau_N^{n,m}$ consistent with
the action of $\langle s_2,T_1,T_2,T_3,T_4\rangle$ is also
consistent with the action of $R_1$ because
\begin{gather*}
 R_1=s_2{T_2}^{-1}.
\end{gather*}
Therefore, we construct $\tau_N^{n,m}$ consistent with
the action of $\langle s_2,T_1,T_2,T_3,T_4\rangle$.
The action of $s_2$ on $\tau_N^{n,m}$ is
\begin{gather}\label{eqn:s2_on_tau}
 s_2(\tau_N^{n,m})=\tau_N^{n-m,-m}.
\end{gather}
We consider only~$\tau_0^{n,m}$ and~$\tau_1^{n,m}$
because $\tau_N^{n,m}$ is determined as a rational function in~$\tau_0^{n,m}$ and~$\tau_1^{n,m}$.
It easily verif\/ied that~$\tau_0^{n,m}$,
(\ref{eqn:qp3_hyper_tau}) (or~(\ref{eqn:gage_qp3_tau0})),
is consistent with the action of $s_2$.
When $N=1$, we rewrite~(\ref{eqn:s2_on_tau}) as
\begin{gather}\label{eqn:s2_Fnm}
 s_2(F_{n,m-1})={\alpha_2}^{-12}Q^{-12m}\,
 \cfrac{\Theta({\alpha_0}^{-12}Q^{12m-12n};Q^{12})}
 {\Theta({\alpha_0}^{-12}{\alpha_2}^{-12}Q^{-12n};Q^{12})}\, F_{n-m,-m-1},
\end{gather}
from (\ref{eqn:qp3_hyper_tau}).
Moreover, by using (\ref{qp3:entry_2}),
(\ref{eqn:s2_Fnm}) can be rewritten as
\begin{gather*}
  \cfrac{s_2(A_{n,m})-B_{n,m}}{({\alpha_2}^{12}Q^{12m};Q^{12})_\infty} \
  {}_1\varphi_1\left(\begin{matrix}0\\ {\alpha_2}^{-12}Q^{-12m+12}\end{matrix}
  ;Q^{12},{\alpha_0}^{12}Q^{12n-12m+12}\right)
  \nonumber\\
 =\cfrac{(A_{n,m}-s_2(B_{n,m}))\Theta({\alpha_0}^{12}Q^{12n-12m};Q^{12})}
  {({\alpha_2}^{-12}Q^{-12m};Q^{12})_\infty
  \Theta({\alpha_0}^{12}{\alpha_2}^{12}Q^{12n};Q^{12})} \
  {}_1\varphi_1\left(\begin{matrix}0\\ {\alpha_2}^{12}Q^{12m+12}\end{matrix}
  ;Q^{12},{\alpha_0}^{12}{\alpha_2}^{12}Q^{12n+12}\right),
\end{gather*}
which implies that $\tau_1^{n,m}$
is also consistent with the action of $s_2$ when
\begin{gather}\label{eqn:condition_s2_tau}
 s_2(A_{n,m})=B_{n,m}.
\end{gather}

\begin{lemma}
Under the assumption \eqref{eqn:condition_s2_tau},
the hypergeometric $\tau$ functions \eqref{eqn:qp3_hyper_tau}
are consistent with the action of $\langle s_2,T_1,T_2,T_3,T_4\rangle$.
\end{lemma}

Therefore we easily obtain the following theorem:

\begin{theorem}\label{theo:qp3_qp2_hyper}
Setting
\begin{gather}
 R_1(A_{n,m})=B_{n,m},\label{eqn:qp3_qp2_hyper}\\
 \alpha_2=Q^{1/2},\nonumber
\end{gather}
and putting
\begin{gather*}
 \tau_N^{2n}=\tau_N^{n,0},\qquad
 \tau_N^{2n-1}=\tau_N^{n,1},
\end{gather*}
we obtain the hypergeometric $\tau$ functions of $q$-{\rm P}$_{\rm II}$.
Here $\tau_N^{n,m}$ is given by \eqref{eqn:qp3_hyper_tau}.
\end{theorem}

In general, the entries of determinants of
the hypergeometric $\tau$ functions of Painlev\'e systems
are expressed by two-parameter family of the functions
satisfying the contiguity relations.
However the hypergeometric $\tau$ functions of $q$-{\rm P}$_{\rm II}$
in Theorem~\ref{theo:qp3_qp2_hyper} have only one parameter
because of the condition~(\ref{eqn:qp3_qp2_hyper}).
In the next section, we construct the hypergeometric $\tau$
functions of $q$-{\rm P}$_{\rm II}$ which admits two parameters.

\subsubsection[Hypergeometric $\tau$ functions of $q$-P$_{\rm II}$ (II)]{Hypergeometric $\boldsymbol{\tau}$ functions of $\boldsymbol{q}$-P$\boldsymbol{{}_{\rm II}}$ (II)}

We construct the hypergeometric $\tau$ functions of $q$-P$_{\rm II}$ whose
ratios correspond to the hypergeometric solutions
of $q$-P$_{\rm II}$ in Proposition~\ref{prop:sol_qp2}.
By the action of the af\/f\/ine Weyl group,
$\tau_N^k$ is determined as a rational function of
$\tau_0^k$ and $\tau_1^k$ (or $\tau_i$ and $\overline{\tau}_i$).
Thus, our purpose is determining~$\tau_0^k$ and~$\tau_1^k$
consistent with the action of $\langle R_1,T_4\rangle$
and constructing $\tau_N^k$ under the conditions
\begin{gather}\label{eqn:qp2_condition_gamma}
 \alpha_2=Q^{1/2},\qquad
 \gamma=1,
\end{gather}
and the boundary condition
\begin{gather}\label{eqn:qp2_condition_tau-1}
 \tau_N^k=0\qquad (N<0).
\end{gather}

First we consider the condition for $\tau_0^k$
which follows from the boundary condition~(\ref{eqn:qp2_condition_tau-1}).
We use the bilinear equation obtained in~\cite{KNT;projective}:

\begin{proposition}
The following bilinear
equation holds:
\begin{gather}\label{eqn:qp2_toda}
 \tau_{N+1}^{k}\tau_{N-1}^{k+1}
 - Q^{(k-4N+1)/2}\gamma^{-2}\alpha_0\tau_{N}^{k+2}\tau_{N}^{k-1}
 - Q^{-k+4N-1}\gamma^{4}{\alpha_0}^{-2}\tau_{N}^{k+1}\tau_{N}^{k} = 0.
\end{gather}
\end{proposition}

By putting $N=0$ in (\ref{eqn:qp2_toda}), we get
\begin{gather}\label{eqn:qp2_toda_2}
 Q^{3(k+1)/2}{\alpha_0}^3\tau_{0}^{k+2}\tau_{0}^{k-1}
 + \tau_{0}^{k+1}\tau_{0}^{k} = 0.
\end{gather}
We set
\begin{gather}\label{eqn:gage_qp2_tau0}
 \tau_{0}^{k} = \Gamma\big(Q^{(2k+3)/2}{\alpha_0}^{2};Q,Q\big)
  \Gamma\big(Q^{-k/2}{\alpha_0}^{-1};Q,Q\big)
  \Gamma\big(Q^{(-k+3)/2}{\alpha_0}^{-1};Q,Q\big)A_1^k.
\end{gather}
From (\ref{eqn:qp2_toda_2}), $A_1^k$ satisf\/ies
\begin{gather}\label{eqn:qp2_tau_0}
 A_{1}^{k+2}A_{1}^{k-1}=A_{1}^{k+1}A_{1}^{k}.
\end{gather}

We next determine $\tau_0^k$ and $\tau_1^k$.
From (\ref{eqn:def_R1}) and Proposition \ref{prop:weyl_group_tau},
we see that the action of $R_1$ on $\tau_0^k$ and $\tau_1^k$ is given by
\begin{gather}
 R_1(\tau_0) = \tau_2,\\
 R_1(\tau_1) =
  \cfrac{Q^{-2}{\alpha_0}^6\tau_{1}\overline{\tau}_{2}+\overline{\tau}_{1}\tau_{2}}
   {Q^{-1}{\alpha_0}^3\overline{\tau}_0},\label{eqn:qp2_act_r1_1}\\
 R_1(\tau_2) = \tau_1,\\
 R_1(\overline{\tau}_0) = \overline{\tau}_2,\\
 R_1(\overline{\tau}_1) =
  \cfrac{Q^{2}{\alpha_0}^6\overline{\tau}_{1}\tau_{2}+\tau_{1}\overline{\tau}_{2}}
   {Q{\alpha_0}^3\tau_0},\label{eqn:qp2_act_r1_2}\\
 R_1(\overline{\tau}_2) = \overline{\tau}_1.\label{eqn:qp2_act_r1_2+}
\end{gather}
From (\ref{eqn:taukN_tau}), we rewrite (\ref{eqn:qp2_act_r1_1})
and (\ref{eqn:qp2_act_r1_2}) as
\begin{gather}
 Q{\alpha_0}^{-3}\tau_1^{-2}\tau_0^1
 - Q^2{\alpha_0}^{-6}\tau_1^0\tau_0^{-1}
 - \tau_0^0\tau_1^{-1} = 0,\label{eqn:qp2_tau_1}\\
 Q^{-1}{\alpha_0}^{-3}\tau_0^{-2}\tau_1^1
 - Q^{-2}{\alpha_0}^{-6}\tau_0^0\tau_1^{-1}
 - \tau_1^0\tau_0^{-1} = 0,\label{eqn:qp2_tau_2}
\end{gather}
respectively.
Setting
\begin{gather}\label{eqn:gage_qp2_tau1}
 \tau^k_{1}=\cfrac{\tau_0^k}{\Theta(Q^{3k+1}{\alpha_0}^{6};Q^3)}\, G_k,
\end{gather}
then the systems (\ref{eqn:qp2_tau_1}) and (\ref{eqn:qp2_tau_2})
reduce to~(\ref{qp2:3-term}).
Therefore $G_k$ is equivalent to~(\ref{qp2:entry}).
If we assume $A_1^k$ is an arbitrary constant,
it does not contradict
(\ref{eqn:qp2_tau_0})--(\ref{eqn:qp2_act_r1_2+}).
Therefore, we may put $A_1^k=1$.

Finally we present an explicit formula for $\tau_N^k$.

\begin{theorem}
Under the assumption \eqref{eqn:qp2_condition_gamma}
and \eqref{eqn:qp2_condition_tau-1},
the hypergeometric $\tau$ functions of $q$-{\rm P}$_{\rm II}$ are given
as the follows:
\begin{gather*}
 \tau_N^k= (-1)^{N(N-1)/2}Q^{N(N-1)(k+N)}{\alpha_0}^{2N(N-1)}\nonumber\\
\phantom{\tau_N^k=}{} \times\cfrac{\Gamma(Q^{(2k+3)/2}{\alpha_0}^{2};Q,Q)
  \Gamma(Q^{-k/2}{\alpha_0}^{-1};Q,Q)
  \Gamma(Q^{(-k+3)/2}{\alpha_0}^{-1};Q,Q)}
  {\Theta(Q^{3k+1}{\alpha_0}^{6};Q^3)^N}\, \phi_N^k,
\end{gather*}
where
\begin{gather*}
 \phi_N^k =
 \begin{vmatrix}
  G_{k}&G_{k-1}&\cdots&G_{k-N+1}\\
  G_{k+2}&G_{k+1}&\cdots&G_{k-N+3}\\
  \vdots&\vdots&\ddots&\vdots\\
  G_{k+2N-2}&G_{k+2N-3}&\cdots&G_{k+N-1}
 \end{vmatrix},\qquad
 \phi_0^k=1,\qquad \phi_{-N}^k=0 \quad (N>0),
\end{gather*}
and
\begin{gather*}
 G_k=
 A_k\Theta\big(ia_0q^{(2k+1)/4};q^{1/2}\big) \
  {}_1\varphi_1\left(\begin{matrix}0\\-q^{1/2}\end{matrix}
  ;q^{1/2},-ia_0q^{(3+2k)/4}\right)\\
\phantom{G_k=}{} +B_k\Theta\big({-}ia_0q^{(2k+1)/4};q^{1/2}\big) \
  {}_1\varphi_1\left(\begin{matrix}0\\-q^{1/2}\end{matrix}
  ;q^{1/2},ia_0q^{(3+2k)/4}\right).
\end{gather*}
Here, $A_k$ and $B_k$ are periodic functions of period one.
\end{theorem}

\begin{proof}
We set
\begin{gather*}
 \tau_N^k= (-1)^{N(N-1)/2}Q^{N(N-1)(k+N)}{\alpha_0}^{2N(N-1)}\\
\phantom{\tau_N^k=}{} \times\cfrac{\Gamma(Q^{(2k+3)/2}{\alpha_0}^{2};Q,Q)
  \Gamma(Q^{-k/2}{\alpha_0}^{-1};Q,Q)
  \Gamma(Q^{(-k+3)/2}{\alpha_0}^{-1};Q,Q)}
  {\Theta(Q^{3k+1}{\alpha_0}^{6};Q^3)^N}\,\phi_N^k.
\end{gather*}
From (\ref{eqn:qp2_condition_tau-1}),
(\ref{eqn:gage_qp2_tau0}), and (\ref{eqn:gage_qp2_tau1}),
we f\/ind that
\begin{gather*}
 \phi_N^k=0 \quad (N<0),\qquad
 \phi_0^k=1,\qquad
 \phi_1^k=G_k.
\end{gather*}
From (\ref{eqn:qp2_toda}), $\phi_N^k$ satisf\/ies
\begin{gather}
  \phi^{k}_{N+1}\phi^{k+1}_{N-1}
  -\phi^{k}_{N}\phi^{k+1}_{N}
  +\phi^{k+2}_{N}\phi^{k-1}_{N} = 0,\label{eqn:qp2_dToda}
\end{gather}
which is a variant of the discrete Toda equation. Under the conditions
\begin{gather*}
 \phi^{k}_N=0\quad (N<0),\qquad
 \phi^{k}_0=1,\qquad
 \phi^{k}_1=c_k,
\end{gather*}
where $c_k$ is an arbitrary function.
Equation~(\ref{eqn:qp2_dToda}) admits a solution expressed by
\begin{gather*}
 \phi^{k}_N = \det\left(c_{k+2i-j-1}\right)_{i,j=1,\ldots,N}\quad (N>0).
\end{gather*}
This complete the proof.
\end{proof}

\subsection[Relation between the hypergeometric $\tau$ functions of $q$-P$_{\rm III}$ and Heine's transform]{Relation between the hypergeometric $\boldsymbol{\tau}$ functions\\ of $\boldsymbol{q}$-P$\boldsymbol{{}_{\rm III}}$ and Heine's transform}

Masuda showed that the consistency of a certain ref\/lection transformation to
the hypergeometric $\tau$~functions of type $E_8^{(1)}$
correspond to Bailey's four term transformation formula~\cite{Masuda:E8}.
It is also shown that the consistency of a certain ref\/lection transformation to
the hypergeometric $\tau$~functions of type~$E_7^{(1)}$ correspond to
limiting case of Bailey's~${}_{10}\varphi_9$ transformation formula~\cite{Masuda:E7}.
We here show that the consistency of $s_0$ to
the hypergeometric $\tau$ functions of $q$-P$_{\rm III}$
give rise to a transformation of~${}_1\varphi_1$
which is obtained by Heine's transform for~${}_2\varphi_1$.

The action of $s_0$ on $\tau_N^{n,m}$ is
\begin{gather}\label{eqn:s0_on_tau}
 s_0(\tau_N^{n,m})=\tau_N^{-n,m-n}.
\end{gather}
We consider only $\tau_0^{n,m}$ and $\tau_1^{n,m}$
because $\tau_N^{n,m}$ is determined as a rational function in~$\tau_0^{n,m}$ and~$\tau_1^{n,m}$.
It easily verif\/ied that $\tau_0^{n,m}$,
(\ref{eqn:qp3_hyper_tau}) (or~(\ref{eqn:gage_qp3_tau0})),
is consistent with the action of $s_0$.
When $N=1$, (\ref{eqn:s0_on_tau}) implies
\begin{gather}\label{eqn:s0_Fnm}
 s_0(F_{n,m-1})=
 \cfrac{\Theta({\alpha_2}^{-12}Q^{-12n+12m};Q^{12})}
 {\Theta({\alpha_0}^{-12}{\alpha_2}^{-12}Q^{12m};Q^{12})}\,F_{-n,m-n-1},
\end{gather}
from (\ref{eqn:qp3_hyper_tau}).
Moreover, by using (\ref{qp3:entry_2}),
(\ref{eqn:s0_Fnm}) can be rewritten as
\begin{gather}
  s_0(A_{n,m})
  {}_1\varphi_1\left(\begin{matrix}0\\{\alpha_0}^{12}{\alpha_2}^{12}Q^{-12m+12}\end{matrix}
  ;Q^{12},{\alpha_2}^{12}Q^{12n-12m+12}\right)\nonumber\\
  \qquad{}-A_{n,m}\cfrac{({\alpha_2}^{12}Q^{12n-12m+12};Q^{12})_\infty}
  {({\alpha_0}^{12}{\alpha_2}^{12}Q^{-12m+12};Q^{12})_\infty} \
  {}_1\varphi_1\left(\begin{matrix}0\\ {\alpha_2}^{12}Q^{12n-12m+12}\end{matrix}
  ;Q^{12},{\alpha_2}^{12}{\alpha_0}^{12}Q^{-12m+12}\right)\nonumber\\
  \qquad{} -B_{n,m}{\alpha_0}^{12}Q^{-12n}\,
  \cfrac{({\alpha_0}^{-12}{\alpha_2}^{-12}Q^{12m},{\alpha_2}^{-12}Q^{-12n+12m+12};Q^{12})_\infty}
  {\Theta({\alpha_0}^{12}Q^{-12n};Q^{12})}\nonumber\\
  \qquad{}\times \;
  {}_1\varphi_1\left(\begin{matrix}0\\ {\alpha_2}^{-12}Q^{-12n+12m+12}\end{matrix}
  ;Q^{12},{\alpha_0}^{12}Q^{-12n+12}\right)\nonumber\\
 \qquad{}
  +s_0(B_{n,m})\,
  \cfrac{({\alpha_0}^{-12}{\alpha_2}^{-12}Q^{12m};Q^{12})_\infty
  \Theta({\alpha_2}^{12}Q^{12n-12m};Q^{12})}
  {({\alpha_0}^{12}{\alpha_2}^{12}Q^{-12m};Q^{12})_\infty
  \Theta({\alpha_0}^{-12}Q^{12n};Q^{12})} \nonumber\\
  \qquad{}\times \;
  {}_1\varphi_1\left(\begin{matrix}0\\{\alpha_0}^{-12}{\alpha_2}^{-12}Q^{12m+12}\end{matrix}
  ;Q^{12},{\alpha_0}^{-12}Q^{12n+12}\right)=0.\label{eqn:s0_Fnm_2}
\end{gather}
In particular, setting
\begin{gather*}
 s_0(A_{n,m})=A_{n,m},\qquad B_{n,m}=0,
\end{gather*}
in (\ref{eqn:s0_Fnm_2}), we obtain
\begin{gather}
  {}_1\varphi_1\left(\begin{matrix}0\\{\alpha_0}^{12}{\alpha_2}^{12}Q^{-12m+12}\end{matrix}
  ;Q^{12},{\alpha_2}^{12}Q^{12n-12m+12}\right)\label{eqn:s0_Fnm_3}\\
 \qquad{} =\cfrac{({\alpha_2}^{12}Q^{12n-12m+12};Q^{12})_\infty}
  {({\alpha_0}^{12}{\alpha_2}^{12}Q^{-12m+12};Q^{12})_\infty} \
  {}_1\varphi_1\left(\begin{matrix}0\\ {\alpha_2}^{12}Q^{12n-12m+12}\end{matrix}
  ;Q^{12},{\alpha_2}^{12}{\alpha_0}^{12}Q^{-12m+12}\right).\nonumber
\end{gather}
Equation (\ref{eqn:s0_Fnm_3}) corresponds to
a specialization of Heine's transform.
Actually, by putting
\begin{gather*}
 a=b^{-1}c,\qquad d=b^{-1}z,
\end{gather*}
in Heine's transform~\cite{ZW:heine}
\begin{gather*}
 {}_2\varphi_1\begin{pmatrix}\begin{matrix}a,b\\c\end{matrix};q,d\end{pmatrix}
 =\cfrac{(a,bd;q)_\infty}{(c,d;q)_\infty} \
  {}_2\varphi_1\begin{pmatrix}\begin{matrix}a^{-1}c,
  d\\bd\end{matrix};q,a\end{pmatrix},
\end{gather*}
we obtain
\begin{gather}\label{eqn:2phi1_heine}
 {}_2\varphi_1\begin{pmatrix}\begin{matrix}b^{-1}c,b\\c\end{matrix}
  ;q,b^{-1}z\end{pmatrix}
 =\cfrac{(b^{-1}c,z;q)_\infty}{(c,b^{-1}z;q)_\infty} \
  {}_2\varphi_1\begin{pmatrix}\begin{matrix}b,b^{-1}z\\z\end{matrix}
  ;q,b^{-1}c\end{pmatrix}.
\end{gather}
Taking the limit $b\to \infty$ in
(\ref{eqn:2phi1_heine}) leads to
\begin{gather*}
 {}_1\varphi_1\begin{pmatrix}\begin{matrix}0\\c\end{matrix};q,z\end{pmatrix}
 =\cfrac{(z;q)_\infty}{(c;q)_\infty} \
  {}_1\varphi_1\begin{pmatrix}\begin{matrix}0\\z\end{matrix};q,c\end{pmatrix},
\end{gather*}
which is equivalent to~(\ref{eqn:s0_Fnm_3}).

\subsection*{Acknowledgements}

The author would like to express sincere thanks to
Professor T.~Masuda for fruitful discussions and valuable suggestions.
I acknowledge continuous
encouragement by Professors K.~Kajiwara and  T.~Tsuda.
This work has been partially supported by the JSPS Research Fellowship.

\pdfbookmark[1]{References}{ref}
\LastPageEnding

\end{document}